\documentstyle[seceq,subeqn,twocolumn]{jpsj}
\def\i{\mbox{\rm i}}
\def\Define{\mathop{\stackrel{\rm def}{=}}}
\def\Bruhat{\mathop{\stackrel{\rm B}{<}}}
\def\dominance{\mathop{\stackrel{\rm D}{<}}}
\def\mae{\hspace{3mm}}

\def\haba{\hspace{6mm}}
\newcommand{\myref}[1]{(\ref{#1})}
\def\nn{\nonumber}

\def\runtitle{Rodrigues Formula for the Nonsymmetric Multivariable
Laguerre Polynomial}
\def\runauthor{Akinori {\sc Nishino}, 
Hideaki {\sc Ujino} and Miki {\sc Wadati}}

\title{
Rodrigues Formula for the Nonsymmetric Multivariable Laguerre
Polynomial
}
\author{
Akinori {\sc Nishino}\footnote{E-mail address: %
nishino@monet.phys.s.u-tokyo.ac.jp},
Hideaki {\sc Ujino}\footnote{E-mail address: %
ujino@monet.phys.s.u-tokyo.ac.jp}
and Miki {\sc Wadati}
\\
}
\inst{
Department of Physics, Graduate School of Science,
University of Tokyo,\\
Hongo 7--3--1, Bunkyo-ku, Tokyo 113--0033
}
\recdate{\hspace{3cm}}
\abst{
Extending a method developed by Takamura and Takano, 
we present the Rodrigues formula for the nonsymmetric 
multivariable Laguerre polynomials which form the orthogonal basis 
for the $B_{N}$-type Calogero model
with distinguishable particles. 
Our construction makes it possible for the first time
to algebraically generate all the nonsymmetric
multivariable Laguerre polynomials
with different parities for each variable.
}
\kword{
nonsymmetric multivariable Laguerre polynomial, 
$B_{N}$-Calogero model, Dunkl-Cherednik operator, Rodrigues formula
}

\begin{document}
\sloppy
\maketitle

%%%%%%%%%%%%%%%%%%%%%%%%%%%%%%%%%%%%%%%%%%%%%%%%%%%%%%%%%%%%%%%%%%%%%%%%%%%%%
\section{Introduction}
Quantum many-body problems have played an important role 
in various fields of physics. In particular, exactly solvable models
have been extensively studied by many physicists and mathematicians.
Since exactly solvable models in quantum mechanics have the same number
of mutually commuting conserved operators as their degrees of freedom,
they are usually called quantum integrable systems.
A good example of them is the Calogero-Sutherland type 
model (C-S model)~\cite{Calogero,Sutherland_1,Sutherland_2}
which has inverse-square long-range interactions between particles
in a one-dimensional space. 
The Dunkl-Cherednik operator formulation is a very powerful tool
for systematic construction of the conserved operators for 
many kinds of C-S models~\cite{Dunkl,Cherednik}.
Their simultaneous eigenfunctions, which form the orthogonal bases
of their Hilbert spaces, are written by products of
the Jastrow type ground state wave function and 
multivariable orthogonal polynomials that are generalized versions
of classical orthogonal polynomials.

The Sutherland model describing the particles on a circle
is a representative of the family of the C-S models.
The orthogonal basis for the Sutherland model has been 
known as the Jack 
polynomial~\cite{Jack,Macdonald_1,Lapointe_1,Lapointe_2}. 
Richness of its analytic properties 
enables us to calculate the Green function, 
the density-density correlation function and so
on~\cite{Ha_1,Ha_2,Takemura_1,Takemura_2}.
That is why studies on the multivariable orthogonal polynomials
have interested many physicists.

On the other hand, 
the Calogero model describing the particles 
in the harmonic potential~\cite{Calogero}
has been treated by the Dunkl-Cherednik operator formulation
as well as the Sutherland model.
The orthogonal basis of the Calogero model is 
a one-parameter deformation of the Jack polynomial,
i.e.\ the Hi-Jack polynomial~\cite{Ujino_1,Ujino_2},
which is called the multivariable Hermite polynomial
from the viewpoint of generalization of the classical 
orthogonal polynomial.
The orthogonal bases for other C-S models were also
constructed~\cite{Baker_1}.

Under the stimulus of the Haldane-Shastry model~\cite{Haldane_1,Shastry},
the C-S models with spin degrees of freedom 
appeared~\cite{Haldane_2,Bernard_1,Bernard_2,Hikami}.
To investigate the orbital part of the eigenfunction,
the C-S models with distinguishable particles were introduced.
The Hamiltonians for these C-S models have
the coordinate exchange operator $K_{jk}$ acting on an arbitrary 
$N$-variable function as follows, 
$$
  K_{jk}f(\cdots,x_{j},\cdots,x_{k},\cdots)
  =f(\cdots,x_{k},\cdots,x_{j},\cdots).
$$
To obtain eigenfunctions of
the C-S models with distinguishable particles, 
it is enough to consider nonsymmetric polynomials 
diagonalizing their Cherednik operators.
Knop and Sahi recursively constructed the nonsymmetric Jack 
polynomial for the Sutherland model 
with distinguishable particles~\cite{Knop}. 
Baker and Forrester translated it to other nonsymmetric 
polynomials~\cite{Baker_2,Baker_3}.

Recently, 
Takamura and Takano algebraically constructed the nonsymmetric Jack 
polynomial~\cite{Takamura}.
The essential point of their work is
introduction of two types of operators, i.e.\
the braid-exclusion and the raising operators.
In the previous paper~\cite{Ujino_3}, we considered 
the Calogero model with distinguishable particles, 
\begin{equation} 
  \label{eq:A-Calogero_1} 
  \mae\hat{H}_{\rm C}^{(A)}
  =\frac{1}{2}\sum_{j=1}^{N}(p_{j}^{2}+\omega^{2}x_{j}^{2}) 
  +\frac{1}{2}\sum_{\stackrel{\scriptstyle j,k=1}{j\neq k}}^{N}
  \frac{a(a-K_{jk})}{(x_{j}-x_{k})^{2}},
\end{equation}
where the constants $a$ and $\omega$ are the coupling parameter 
and the strength of the external harmonic well, respectively,
and $p_{j}=\frac{1}{\i}\frac{\partial}{\partial x_{j}}$.
Generalizing the method developed by Takamura and Takano, 
we provided 
the Rodrigues formula for the nonsymmetric
multivariable Hermite polynomial.
While it is difficult to compute the coefficient of 
the top term of the polynomial,
it is easy to give simple expressions for the nonsymmetric
eigenfunctions and to calculate their norms
by the Takamura-Takano method.

The above Calogero Hamiltonian~\myref{eq:A-Calogero_1} 
is invariant under the action of the  $A_{N-1}$-type 
Weyl group, i.e.\ under $S_{N}$, on the indices of the particle.
Thus the model is sometimes called the $A_{N-1}$-Calogero model.
Similarly,
the Calogero models associated with other Weyl groups exist.
The system, which has the additional interactions 
with the boundary at the origin
and the mirror image particles, is called the $B_{N}$-Calogero 
model~\cite{Yamamoto,Kakei}.
The orthogonal basis for the $B_{N}$-Calogero model is
known to be the multivariable Laguerre polynomial which is
even in each variable.
For the $B_{N}$-Calogero model with distinguishable particles,
i.e.\ including both the coordinate exchange and reflection operators,
the nonsymmetric eigenfunctions of the Cherednik operators 
have already been constructed by the Knop-Sahi method~\cite{Baker_3}.
However they are also restricted to functions which are even
in each variable.
In this paper, extending the Takamura-Takano method, 
we shall present the Rodrigues formula 
for the nonsymmetric multivariable Laguerre polynomial.
Our method enables an algebraic construction of the
nonsymmetric multivariable Laguerre polynomial which
holds different parities for each variable.

The plan of the paper is the following. 
Section \ref{sec:review} is devoted to a summary of the
Dunkl-Cherednik operator formulation for the $B_{N}$-Calogero model.
In \S\ref{sec:Rodrigues}, we shall present the Rodrigues formula 
for the nonsymmetric multivariable Laguerre polynomial
which is allowed to hold different parities for each variable.
We shall calculate their norms in \S\ref{sec:norm}.
The final section is devoted to concluding remarks.

%%%%%%%%%%%%%%%%%%%%%%%%%%%%%%%%%%%%%%%%%%%%%%%%%%%%%%%%%%%%%%%%%%%%%%%%%%%%
\section{Dunkl-Cherednik Operator Formulation for the 
$B_{N}$-Calogero Model}
\label{sec:review}
We briefly summarize the Dunkl-Cherednik operator formulation
for the $B_{N}$-Calogero model.
The $B_{N}$-Calogero Hamiltonian with both the coordinate exchange and 
reflection operators is defined by
\begin{eqnarray}
  \label{eq:B-Calogero_1}
  \hspace{-4mm}\hat{H}_{\rm C}^{(B)} 
  &=&\frac{1}{2}\sum_{j=1}^{N}
  \left(p_{j}^{2}+\omega^{2}x_{j}^{2}
  +\frac{b(b-t_{j})}{x_{j}^{2}}\right) \nn \\ 
  &+&\frac{1}{2}\sum_{\stackrel{\scriptstyle j,k=1}{j\neq k}}^{N}
  \left(\frac{a(a-K_{jk})}{(x_{j}-x_{k})^{2}}
  +\frac{a(a-t_{j}t_{k}K_{jk})}{(x_{j}+x_{k})^{2}}\right),
\end{eqnarray}
where $b$ is another coupling parameter
and $t_{j}$ is the reflection operator,
\begin{equation}
  \label{eq:reflection}
  t_{j}f(\cdots,x_{j},\cdots)
  =f(\cdots,-x_{j},\cdots).
\end{equation}
Hereafter we call eq.~\myref{eq:B-Calogero_1}
the $B_{N}$-Calogero model with distinguishable particles.
The ground state wave function and the ground state 
energy are
\begin{eqnarray}
  \label{eq:B-gs}
  &&\phi^{(B)}_{\rm g}({\mib x})
  =\!\!\!\prod_{1\le j<k\le N}\!\!\!\!
   |x_{j}^{2}-x_{k}^{2}|^{a}
   \prod_{l=1}^{N}|x_{l}|^{b}
   \exp\Big(\!\!
   -\frac{1}{2}\omega\!\sum_{m=1}^{N}x_{m}^{2}\Big),
  \nn\\ 
  &&E_{\rm g}^{(B)}=\frac{1}{2}\omega N(1+2(N-1)a+2b),
\end{eqnarray}
respectively.
For the sake of simplicity, we write
$f(\mib{x})$ for a function with $N$ variables,
$f(x_{1},x_{2},\cdots,x_{N})$.
To investigate the polynomial part of the eigenfunction,
we carry out a similarity transformation on the 
Hamiltonian~\myref{eq:B-Calogero_1}
with the ground state~\myref{eq:B-gs},
\begin{eqnarray}
  \label{eq:B-Calogero_2}
  H_{\rm C}^{(B)}
  &\Define&
     (\phi_{\rm g}^{(B)}({\mib x}))^{-1}
     (\hat{H}_{\rm C}^{(B)}-E_{\rm g}^{(B)})
     \phi_{\rm g}^{(B)}({\mib x}) \nn\\
  &=&\sum_{j=1}^{N}\left(
     \omega x_{j}\frac{\partial}{\partial x_{j}}
     +\frac{1}{2}\frac{\partial^{2}}{\partial x_{j}^{2}}
     +\frac{b}{2 x_{j}}\frac{\partial}{\partial x_{j}}
     +\frac{b(t_{j}-1)}{2 x_{j}^{2}}\right)  \nn\\
  & &+\frac{1}{2} a\sum_{j\neq k}\left(
     \frac{2}{x_{j}^{2}-x_{k}^{2}}
     \Big(x_{j}\frac{\partial}{\partial x_{j}}
         -x_{k}\frac{\partial}{\partial x_{k}}\Big)\right. \nn\\
  & &\hspace{20mm}\left.
     +\frac{K_{jk}-1}{(x_{j}-x_{k})^{2}}
     +\frac{t_{j}t_{k}K_{jk}-1}{(x_{j}+x_{k})^{2}}\right).
\end{eqnarray}
For the transformed Hamiltonian, 
the ground state wave function is unity
and the ground state energy is zero.
In what follows, we sometimes call the operator~\myref{eq:B-Calogero_2}
the $B_{N}$-Calogero Hamiltonian.

The Dunkl operators associated with the $B_{N}$-type Weyl group
are defined as
\begin{eqnarray}
  \label{eq:B-Dunkl}
  \nabla_{j}
  &\Define&\frac{\partial}{\partial x_{j}}
  +\frac{b(1-t_{j})}{x_{j}}  \nn\\
  & &+a\!\!\sum_{k(\neq j)}
    \left(\frac{1-K_{jk}}{x_{j}-x_{k}}
         +\frac{1-t_{j}t_{k}K_{jk}}{x_{j}+x_{k}}\right).
\end{eqnarray}
The commutation relation with coordinates $x_{j}$ is given by
\begin{eqnarray}
  \big[\nabla_{j},x_{k}\big]
  &=&\delta_{jk}
  \Big(1+a\!\!\sum_{l(\neq j)}(1+t_{j}t_{l})K_{jl}
       +2b\,t_{j}\Big) \nn\\
  & &-(1-\delta_{jk})a(1-t_{j}t_{k})K_{jk}.
\end{eqnarray}
We introduce the creation- and annihilation-like operators 
and Cherednik operators for the $B_{N}$-Calogero model,
\begin{eqnarray}
  &&\alpha_{j}^{\dagger}\Define x_{j}-\frac{1}{2\omega}\nabla_{j},
  \haba
  \alpha_{j}\Define \nabla_{j}, 
  \label{eq:alpha-op} \\
  &&d_{j}\Define\alpha_{j}^{\dagger}\alpha_{j}
    +a\!\!\sum_{k=j+1}^{N}\!(1+t_{j}t_{k})K_{jk}+b\,t_{j}.
  \label{eq:B-Cherednik}
\end{eqnarray}
Strictly,
the operators $\alpha_{j}^{\dagger}$ and $\alpha_{j}$ are 
not Hermitian conjugate each other.
But we still use the notation $\dagger$ 
since they have the relation
$\phi_{\rm g}^{(B)}\alpha_{j}^{\dagger}(\phi_{\rm g}^{(B)})^{-1}
 =\big(\phi_{\rm g}^{(B)}
       \alpha_{j}(\phi_{\rm g}^{(B)})^{-1}\big)^{\dagger}$
after recovering the effect of the ground state~\myref{eq:B-gs}.
Although there exist different definitions for the Cherednik 
operator~\cite{Yamamoto,Kakei},
we select the one~\myref{eq:B-Cherednik}
for convenience of later discussions.
Commutation relations among these Dunkl-Cherednik operators are
given as follows,
\begin{eqnarray}
  &&\big[ \alpha_{j}^{\dagger},\alpha_{k}^{\dagger}\big]
    =\big[\alpha_{j},\alpha_{k} \big]=\big[d_{j},d_{k}\big]=0,  \nn\\
  &&\big[\alpha_{j},\alpha_{k}^{\dagger}\big]
    =\delta_{jk}
    \Big(1+a\sum_{l(\neq j)}
     (1+t_{j}t_{l})K_{jl}+2b\,t_{j}\Big) \nn\\
  &&\hspace{20mm}-(1-\delta_{jk})a(1-t_{j}t_{k})K_{jk}, \nn\\
  &&\big[d_{j},\alpha_{k}^{\dagger}\big] \nn\\
  &&=\!\delta_{jk}\Big(
   \alpha_{j}^{\dagger}\!
   +a\!\sum_{l=1}^{j-1}\!\alpha_{j}^{\dagger}
    (1\!+t_{j}t_{l})K_{jl}\!
   +a\!\!\!\sum_{l=j+1}^{N}\!\!\!\alpha_{l}^{\dagger}
    (1\!-t_{j}t_{l})K_{jl}\!\Big)
   \nn\\
  &&-a\Big(\!\Theta(j\!-\!k)\alpha_{j}^{\dagger}
            (1\!-t_{j}t_{k})K_{jk}\!
           +\!\Theta(k\!-\!j)\alpha_{k}^{\dagger}
            (1\!+t_{j}t_{k})K_{jk}\!\Big),
   \nn\\
  &&\big[d_{j},\alpha_{k}\big]  \nn\\
  &&=\!-\delta_{jk}\Big(
     \alpha_{j}\!
     +a\!\sum_{l=1}^{j-1}
      (1\!+t_{j}t_{l})K_{jl}\alpha_{j}\!
     +a\!\!\!\sum_{l=j+1}^{N}
      \!\!(1\!-t_{j}t_{l})K_{jl}\alpha_{l}\!\Big)
  \nn\\
  &&+a\Big(\!\Theta(j\!-\!k)
            (1\!-t_{j}t_{k})K_{jk}\alpha_{j}\!
          +\!\Theta(k\!-\!j)
            (1\!+t_{j}t_{k})K_{jk}\alpha_{k}\!\Big),
   \nn\\
  &&
\end{eqnarray}
where $\Theta(x)$ is the Heaviside function,
\begin{eqnarray}
  \Theta(x)=\left\{
    \begin{array}{ll}
      0, & (x\leq 0), \\
      1, & (x > 0).
    \end{array}\right. \nn
\end{eqnarray}
The Cherednik operators obey the following relations with
the coordinate exchange and reflection operators,
\begin{eqnarray}
  \label{eq:property-B-Cherednik}
  &&d_{j}K_{j,j+1}-K_{j,j+1}d_{j+1}=a(1+t_{j}t_{j+1}), \nn\\
  &&d_{j+1}K_{j,j+1}-K_{j,j+1}d_{j}=-a(1+t_{j}t_{j+1}), \nn\\
  &&\big[d_{j},K_{k,k+1}\big]=0,\haba (j\neq k, k+1), \nn\\
  &&\big[d_{j},t_{k}\big]=0.
\end{eqnarray}
In terms of the Cherednik operator $d_{j}$,
the Hamiltonian~\myref{eq:B-Calogero_2} is rewritten as
\begin{equation}
  \label{eq:B-Calogero_3}
  H_{\rm C}^{(B)}
  =\omega\sum_{j=1}^{N}\Big(d_{j}-(N-1)a-b\Big).
\end{equation}
It is well-known that 
the power sums of the Cherednik operators
provide the mutually commuting conserved operators for the
$B_{N}$-Calogero model 
since $\{d_{j}\}$ mutually commute, $\big[d_{j},d_{k}\big]=0$.
Equation~\myref{eq:B-Calogero_3} shows that 
the Hamiltonian~\myref{eq:B-Calogero_2}, which is one of the
conserved operators, is written by the commuting Cherednik operators.
Thus the Cherednik operators $\{d_{j}\}$
themselves are 
the conserved operators for the $B_{N}$-Calogero 
model with distinguishable particles.
Because the Cherednik operators and the reflection operators
mutually commute \myref{eq:property-B-Cherednik},
the simultaneous eigenfunction of the Cherednik operators is,
at the same time, the simultaneous eigenfunction of the
reflection operators.
In other words, the parities for each variable are good quantum
numbers.
As we shall see shortly in the definition of the simultaneous
eigenfunction~\myref{eq:m-Laguerre_1}, this property appears as
a restriction on a parity for a variable.

All the eigenfunctions are labeled by the symbol $\lambda_{\sigma}$
(composition) consisting of a partition $\lambda$ and a distinct 
permutation $\sigma\in S_{N}$.
A partition $\lambda$ is a sequence of $N$ nonnegative 
integers,
\begin{eqnarray}
  &&\lambda\Define
    \{\lambda_{1}\geq\lambda_{2}\geq\cdots\geq\lambda_{N}\geq 0\}. 
  \nn
\end{eqnarray}
Distinct permutations $\sigma$ and $\tau$ must satisfy 
$\lambda_{\sigma (j)}\neq\lambda_{\tau (j)}$ 
for some $j\in\{1,2,\cdots,N\}$.
To define the nonsymmetric multivariable Laguerre polynomials,
we introduce the Bruhat order $\Bruhat$,
\begin{eqnarray}
  \label{eq:Bruhat}
  &&\mu_{\tau}\Bruhat\lambda_{\sigma}\Leftrightarrow\left\{
  \begin{array}{cl}
    1) & \mu \dominance \lambda, \\
    2) & \mbox{ when $\mu=\lambda$ then the first } \\ 
       & \mbox{ non-vanishing difference }          \\
       & \tau(j)-\sigma(j)>0, 
  \end{array}\right.\nn
\end{eqnarray}
where the symbol $\dominance$ is the dominance order,
$$
  \mu\dominance\lambda 
  \Leftrightarrow \mu\neq\lambda,
  |\mu|=|\lambda|
  \mbox{ and }
  \sum_{k=1}^{l}\mu_{k}\leq\sum_{k=1}^{l}\lambda_{k},
$$
for all $l=1,2,\cdots,N$.
A set of indistinct permutations for a partition $\lambda$,
$$
  \{\sigma\}
  \Define\{\tau\in S_{N}|\lambda_{\tau}=\lambda_{\sigma}
  \mbox{ and }\lambda_{\tau}\stackrel{\rm B}{>}\lambda_{\sigma}
  \mbox{ for }\tau\neq\sigma\},
$$
is represented by the permutation $\sigma$ which gives the minimum in
the sense of the Bruhat order $\Bruhat$.
The nonsymmetric multivariable Laguerre polynomial,
$l_{\lambda_{\sigma}}(\mib{x};1/a,1/b,\omega)$,
is the nondegenerate
simultaneous eigenfunction of the Cherednik operators $\{d_{j}\}$ 
with the coefficient of its top term $\mib{x}^{\lambda_{\sigma}}
\Define x_{1}^{\lambda_{\sigma(1)}}x_{2}^{\lambda_{\sigma(2)}}\cdots
x_{N}^{\lambda_{\sigma(N)}}$ conventionally taken to be unity,
\begin{subequations}
\begin{eqnarray}
  &&l_{\lambda_{\sigma}}(\mib{x};1/a,1/b,\omega) 
  =\mib{x}^{\lambda_{\sigma}}
  +\hspace{-16mm}
   \sum_{{\scriptstyle \big(\mu_{\tau}\Bruhat\lambda_{\sigma} \;
          {\rm or }\;|\mu|<|\lambda|\big) }\;\atop
         {\scriptstyle 
          {\rm and}
          \;^{\forall} k,\,
          \mu_{\tau(k)}\equiv\lambda_{\sigma(k)}\!\!\pmod{2} }}
   \hspace{-16mm}
   w_{\lambda_{\sigma}\mu_{\tau}}(a,b,\frac{1}{2\omega})
   \mib{x}^{\mu_{\tau}}, \nn\\
  && \label{eq:m-Laguerre_1}\\
  &&d_{j}l_{\lambda_{\sigma}}(\mib{x};1/a,1/b,\omega) 
  =\bar{\lambda}_{\sigma(j)}
     l_{\lambda_{\sigma}}(\mib{x};1/a,1/b,\omega),
  \label{eq:eigenvalue_1}
\end{eqnarray}
\end{subequations}
where
\begin{eqnarray}
  \bar{\lambda}_{\sigma(j)} 
  &\Define&\lambda_{\sigma(j)}
    +2 a\Big(\#\{1\leq k<j|\lambda_{\sigma(k)}<\lambda_{\sigma(j)}\}
  \nn\\
  & & +\#\{j<k\leq N|\lambda_{\sigma(k)}\leq\lambda_{\sigma(j)}\}
         \Big)+b. \nn
\end{eqnarray}
From eq.~\myref{eq:m-Laguerre_1}, we see that
each monomial, $\mib{x}^{\lambda_{\sigma}}$ and 
$\mib{x}^{\mu_{\tau}}$, in
the nonsymmetric multivariable Laguerre polynomial
has the same parity for a variable.
The energy eigenvalue of the eigenfunction $l_{\lambda_{\sigma}}$
is given by
$$
  H_{\rm C}^{(B)}l_{\lambda_{\sigma}}
  = \sum_{k=1}^{N}\lambda_{\sigma(k)}l_{\lambda_{\sigma}}.
$$
The nonsymmetric multivariable Laguerre polynomials
span the orthogonal basis for the $B_{N}$-Calogero model
with distinguishable particles.

%%%%%%%%%%%%%%%%%%%%%%%%%%%%%%%%%%%%%%%%%%%%%%%%%%%%%%%%%%%%%%%%%%%%%%%%%%%%%
\section{Rodrigues Formula for the Nonsymmetric Multivariable 
Laguerre Polynomial}
\label{sec:Rodrigues}
Generalizing the method developed by Takamura
and Takano for the nonsymmetric Jack polynomial~\cite{Takamura},
we shall present the Rodrigues formula for the nonsymmetric multivariable
Laguerre polynomial.

We introduce two types of operators.
The first type is the braid-exclusion operator for 
the $B_{N}$-Calogero model,
\begin{equation}
  \label{eq:X-op}
  X_{j,j+1}\Define \i\big[d_{j},K_{j,j+1}\big],
\end{equation}
which satisfies the following relations,
\begin{eqnarray}
  &&d_{j}X_{j,j+1}-X_{j,j+1}d_{j+1}=0, \nn\\
  &&d_{j+1}X_{j,j+1}-X_{j,j+1}d_{j}=0, \nn\\
  &&\big[d_{j},X_{k,k+1}\big]=0,\haba (j\neq k,k+1), \nn\\
  &&t_{j}X_{j,j+1}-X_{j,j+1}t_{j+1}=0, \nn\\
  &&t_{j+1}X_{j,j+1}-X_{j,j+1}t_{j}=0, \nn\\
  &&[t_{j},X_{k,k+1}]=0,\haba (j\neq k,k+1).
  \label{eq:property-X-op_1}
\end{eqnarray}
From the above relations, we obtain
\begin{eqnarray}
  &&X_{j,j+1}^{2}
    =(d_{j}-d_{j+1})^{2}-2a^{2}(1+t_{j}t_{j+1}), \nn\\
  &&X_{j,j+1}X_{j+1,j+2}X_{j,j+1}
    =X_{j+1,j+2}X_{j,j+1}X_{j+1,j+2}, \nn\\
  &&[X_{j,j+1},X_{k,k+1}]=0, \haba (|j-k|\geq 2), \nn\\
  &&t_{j}X_{j,j+1}t_{j+1}X_{j,j+1}
    =X_{j,j+1}t_{j+1}X_{j,j+1}t_{j}.
  \label{eq:property-X-op_2}
\end{eqnarray}
The second and fourth equations of \myref{eq:property-X-op_2}
are of the same form as the Yang-Baxter and the reflection relations, 
respectively.
The operator $X_{j,j+1}$ is Hermitian, $X_{j,j+1}^{\dagger}=X_{j,j+1}$,
but has no inverse operator
since there exists an eigenstate of the operator $X_{j,j+1}^{2}$
whose eigenvalue is zero.

The operators of the second type
are the raising and lowering operators.
To construct them,
we first introduce the Knop-Sahi operators~\cite{Knop} 
for the $B_{N}$-Calogero model,
\begin{eqnarray}
  \label{eq:Knop-Sahi}
  &&e^{\dagger}
    \Define K_{N,N-1}\cdots K_{2,1} \alpha_{1}^{\dagger}, \nn\\
  &&e\Define
    (e^{\dagger})^{\dagger}=\alpha_{1} K_{1,2}\cdots K_{N-1,N}. 
\end{eqnarray}
The above definitions~\myref{eq:Knop-Sahi}
are different from those introduced by 
Baker and Forrester~\cite{Baker_3}.
We do not impose the restriction to eigenfunctions such that
they should be
even in each variable $x_{j}$.
The Knop-Sahi operator is related to eqs.~\myref{eq:B-Cherednik},
\myref{eq:X-op} and \myref{eq:reflection},
\begin{eqnarray}
  \label{eq:property-KS}
  &&d_{j}e^{\dagger}=e^{\dagger}d_{j+1},\haba (j=1,\cdots,N-1), \nn\\
  &&d_{N}e^{\dagger}=e^{\dagger}(d_{1}+1), \nn\\
  &&X_{j,j+1}e^{\dagger}=e^{\dagger}X_{j+1,j+2},
   \haba (j=1,\cdots,N-2), \nn\\
  &&X_{N-1,N}(e^{\dagger})^{2}=(e^{\dagger})^{2}X_{1,2}, \nn\\
  &&t_{j}e^{\dagger}=e^{\dagger}t_{j+1}, 
   \haba (j=1,\cdots,N-1), \nn\\
  &&t_{N}e^{\dagger}=-e^{\dagger}t_{1}.
\end{eqnarray}
Next, using the braid-exclusion and Knop-Sahi operators,
we define the constituent operators,
\begin{eqnarray}
  \label{eq:constituent}
  &&b_{j}^{\dagger}\Define X_{j,j+1}\cdots X_{N-1,N}e^{\dagger},\quad
    (j=1,\cdots,N-1),\nn\\
  &&b_{N}^{\dagger}\Define e^{\dagger},\nn\\
  &&b_{j}\Define(b_{j}^{\dagger})^{\dagger}=e X_{N,N-1}\cdots X_{j,j+1},
    \quad (j=1,\cdots,N-1),\nn\\
  &&b_{N}\Define(b_{N}^{\dagger})^{\dagger}=e.
\end{eqnarray}
With the help of eqs.~\myref{eq:property-KS},
we have the relations,
\begin{eqnarray}
  &&d_{j}b_{k}^{\dagger}=\left\{
  \begin{array}{ll}
    b_{k}^{\dagger}d_{j+1},   & (1\leq j\leq k-1), \\
    b_{k}^{\dagger}(d_{1}+1), & (j=k), \\
    b_{k}^{\dagger}d_{j},     & (k+1\leq j\leq N),
  \end{array}\right. \nn\\
  &&\lefteqn{X_{j,j+1}b_{k}^{\dagger}} \nn\\
  &&=\left\{
  \begin{array}{ll}
    b_{k}^{\dagger}X_{j,j+1},   &\hspace{-24mm} (1\leq k\leq j-1), \\
    b_{j+1}^{\dagger}((d_{j+1}-d_{1})^{2}-2a^{2}(1-t_{j+1}t_{1})),  
                                & \\
                                &\hspace{-24mm} (k=j),  \\
    b_{j}^{\dagger},            &\hspace{-24mm} (k=j+1), \\
    b_{k}^{\dagger}X_{j+1,j+2}, &\hspace{-24mm}(j+2\leq k\leq N),
  \end{array}\right. \nn\\
  &&t_{j}b_{k}^{\dagger}=\left\{
  \begin{array}{ll}
    b_{k}^{\dagger}t_{j},   & (1\leq k\leq j-1), \\
    -b_{j}^{\dagger} t_{1},  & (k=j), \\
    b_{k}^{\dagger}t_{j+1},  & (j+1\leq k\leq N),
  \end{array}\right. \nn\\
  &&b_{j}^{\dagger}b_{k}^{\dagger}=\left\{
  \begin{array}{ll}
    b_{k}^{\dagger}b_{j+1}^{\dagger}X_{1,2}, & (j<k), \\
    X_{k-1,k}\cdots X_{j-2,j-1}(b_{j}^{\dagger})^{2}, &(j\geq k).
  \end{array}\right.
\end{eqnarray}
In the end,
we define the raising and lowering operators 
for the $B_{N}$-Calogero model as follows,
\begin{eqnarray}
  a_{j}^{\dagger}\Define(b_{j}^{\dagger})^{j},\haba
  a_{j}\Define(b_{j})^{j}.
  \label{eq:raising}
\end{eqnarray}
The commutation relations with the Cherednik 
operator $d_{j}$
are simply expressed by
\begin{eqnarray}
  \big[d_{j},a_{k}^{\dagger}\big]
  &=&\left\{
    \begin{array} {ll}
      0,               &(j>k),  \\
      a_{k}^{\dagger}, &(j\leq k),
    \end{array}\right. \nn\\
  \big[a_{j}^{\dagger},a_{k}^{\dagger}\big]
  &=&\big[a_{j},a_{k}\big]=0.
\end{eqnarray}
From the above relations, we notice that
the raising operator $a_{k}^{\dagger}$ increases the eigenvalue
of the Cherednik operator $d_{j}\;(j \leq k)$ by one.
Thus the simultaneous eigenfunction for the 
Cherednik operators $\{d_{j}\}$ takes the form,
\begin{eqnarray}
  \label{eq:eigenstate_1}
  \tilde{l}_{\lambda}(\mib{x})\Define 
  (a_{1}^{\dagger})^{\lambda_{1}-\lambda_{2}}
  (a_{2}^{\dagger})^{\lambda_{2}-\lambda_{3}}\!\cdots
  (a_{N}^{\dagger})^{\lambda_{N}} |0\rangle,
\end{eqnarray}
where $|0\rangle=1$ is the ground state, 
and the eigenvalue is
\begin{eqnarray}
  \label{eq:eigenvalue_2}
  d_{j}\tilde{l}_{\lambda}
  =(\lambda_{j}+2(N-j)a+b)\,\tilde{l}_{\lambda}
  =\bar{\lambda}_{j} \tilde{l}_{\lambda}.
\end{eqnarray}
Comparing eq.~\myref{eq:eigenvalue_2} with eq.~\myref{eq:eigenvalue_1},
we identify eq.~\myref{eq:eigenstate_1} as the eigenstate 
with the composition $\lambda_{\rm id}$.

Now, we consider the eigenstates 
with the general composition $\lambda_{\sigma}$.
The first and second relations of eqs.~\myref{eq:property-X-op_1} show
that the braid-exclusion operator $X_{j,j+1}$ acts on an eigenstate
to exchange the eigenvalue of $d_{j}$ ( $d_{j+1}$ ) 
for that of $d_{j+1}$ ( $d_{j}$ ).
For a composition $\lambda_{\sigma}$ where the distinct permutation
$\sigma$ is expressed by the product of transpositions as
$$
  \sigma=(k_{l},k_{l}+1)\cdots(k_{2},k_{2}+1)(k_{1},k_{1}+1),
$$
the Rodrigues formula for the nonsymmetric eigenfunction
with the composition $\lambda_{\sigma}$ is presented as follows,
\begin{eqnarray}
  \label{eq:eigenstate_2}
  \!\!\!\tilde{l}_{\lambda_{\sigma}}(\mib{x})
  &\Define& 
    X_{k_{1},k_{1}+1}X_{k_{2},k_{2}+1}\cdots X_{k_{l},k_{l}+1} \nn\\
  &&\times
     (a_{1}^{\dagger})^{\lambda_{1}-\lambda_{2}}
     (a_{2}^{\dagger})^{\lambda_{2}-\lambda_{3}}\!\cdots
     (a_{N}^{\dagger})^{\lambda_{N}} |0\rangle, 
\end{eqnarray}
and the eigenvalue is
\begin{equation}
  d_{j}\tilde{l}_{\lambda_{\sigma}}
    =\bar{\lambda}_{\sigma (j)} \tilde{l}_{\lambda_{\sigma}}.
\end{equation}
Since the nondegenerate eigenfunction $\tilde{l}_{\lambda_{\sigma}}$
have the same eigenvalue of the nonsymmetric multivariable
Laguerre polynomial $l_{\lambda_{\sigma}}$~\myref{eq:m-Laguerre_1}, 
we conclude that $\tilde{l}_{\lambda_{\sigma}}(\mib{x})$
is identified with $l_{\lambda_{\sigma}}(\mib{x})$
up to normalization,
$\tilde{l}_{\lambda_{\sigma}}(\mib{x})\propto
 l_{\lambda_{\sigma}}(\mib{x})$.

We note that the eigenstate~\myref{eq:eigenstate_2} 
is also the simultaneous eigenstate 
of the reflection operators $\{t_{j}\}$
since the Cherednik operator $d_{j}$ commutes with 
all $t_{k}$, i.e.\ $\big[d_{j},t_{k}\big]=0$.
So the parity for a variable $x_{j}$ 
of the nonsymmetric multivariable Laguerre polynomial
is restricted to even or odd
as we have explained in the previous section.
It is possible that the parity for a variable $x_{j}$ 
is different from the one for a variable $x_{k} \,(k\neq j)$.
This means that the eigenstate we have constructed 
is allowed to be the functions whose parities
are different for each variable.
After symmetrization on the variables,
the eigenstate including different parities
for each variable vanishes
and the eigenstate with the single parity for all variables is left.
The result reproduces the one by
Baker and Forrester~\cite{Baker_3}.

%%%%%%%%%%%%%%%%%%%%%%%%%%%%%%%%%%%%%%%%%%%%%%%%%%%%%%%%%%%%%%%%%%%%%%%%%%%%
\section{Norm Formula}
\label{sec:norm}
In this section we shall calculate the norm of the eigenfunction
obtained in the previous section. 
The Knop-Sahi operators for the $B_{N}$-Calogero model
have the following properties,
\begin{eqnarray}
  e^{\dagger} e=d_{N}-b t_{N},\haba
  e e^{\dagger}=d_{1}+b t_{1}+1.  \nn
\end{eqnarray}
From the definition~\myref{eq:constituent}
and the above relations, it is easy to verify
\begin{eqnarray}
  &&b_{j}^{\dagger}b_{j}\!=\!(d_{j}\!-\!bt_{j})\!\!\!
    \prod_{k=j+1}^{N}\!\!\!\!
    \big( (d_{j}\!-\!d_{k})^{2}\!-\!2a^{2}(1\!+\!t_{j}t_{k}) \big), \nn\\
  &&b_{j}b_{j}^{\dagger}\!=\!(d_{1}\!+\!bt_{j}\!+\!1)\!\!\!\!
    \prod_{k=j+1}^{N}\!\!\!\!
    \big( 
    (d_{k}\!-\!d_{1}\!-\!1)^{2}\!-\!2a^{2}\!(1\!-\!t_{k}t_{1})
    \big). \nn
\end{eqnarray}
Owing to the definition of the raising and lowering 
operators~\myref{eq:raising},
we have the following expressions of the number-like operator,
\begin{eqnarray}
  &&a_{j}^{\dagger}a_{j}\!
    =\!\prod_{k=1}^{j}(d_{k}\!-\!bt_{k})\!\!\!
    \prod_{l=j+1}^{N}\!\!\!\!
    \big( 
    (d_{k}\!-\!d_{l})^{2}\!-\!2a^{2}(1\!+\!t_{k}t_{l}) 
    \big), \nn\\
  &&a_{j}a_{j}^{\dagger}\!
    =\!\prod_{k=1}^{j}(d_{k}\!+\!bt_{k}\!+\!1)\!\!\!\!
    \prod_{l=j+1}^{N}\!\!\!\!
    \big( 
    (d_{k}\!-\!d_{l}\!+\!1)^{2}\!-\!2a^{2}(1\!-\!t_{k}t_{l}) 
    \big). \nn
\end{eqnarray}
We recall the definitions of the inner product and the norm
for the $B_{N}$-Calogero model,
\begin{eqnarray}
  \langle f,g\rangle 
  &\Define&\int_{-\infty}^{\infty}
  \prod_{j=1}^{N}{\rm d}x_{j}
  |\phi_{\rm g}^{(B)}|^{2}f^{\dagger}(\mib{x})g(\mib{x}), \nn\\
  |f|^{2} & \Define & \langle f,f\rangle.
\end{eqnarray}
Since the Cherednik operators $d_{j}$ are Hermitian
concerning the inner product,
the nonsymmetric multivariable Laguerre polynomials are orthogonal
with respect to the inner product,
\[
  \langle l_{\lambda_{\sigma}},l_{\mu_{\tau}}\rangle
  =|l_{\lambda_{\sigma}}|^{2}
  \delta_{\lambda_{\sigma},\mu_{\tau}}
  \Leftrightarrow
  \langle \tilde{l}_{\lambda_{\sigma}},
          \tilde{l}_{\mu_{\tau}}\rangle
  =|\tilde{l}_{\lambda_{\sigma}}|^{2}
  \delta_{\lambda_{\sigma},\mu_{\tau}}.
\]
The norm of the eigenfunction 
$\tilde{l}_{\lambda}$ with the composition
$\lambda=\lambda_{\rm id}$ is calculated as follows,
\begin{eqnarray}
  \label{eq:norm-Laguerre_1}
  &&\frac{
    \langle \tilde{l}_{\lambda},\tilde{l}_{\lambda} \rangle
    }{ \langle 1,1 \rangle}
    =\prod_{l=1}^{N}\prod_{m=1}^{l}\!\!
     \prod_{r=1}^{\lambda_{l}-\lambda_{l+1}}\!\!\!\!
     \big[ 
     \bar{\lambda}_{m}\!+\!(-1)^{\lambda_{m}\!-r}b\!-\!r\!+\!1 
     \big] \nn\\
  & &\hspace{20mm}\times\!\!\!\!\prod_{k=l+1}^{N}\!\!\!
     \big[ 
     (\bar{\lambda}_{m}\!-\!\bar{\lambda}_{k}\!-\!r\!+\!1)^{2} \nn\\
  & &\hspace{34mm}
     \!-\!2a^{2}(1\!-\!(-1)^{\lambda_{m}\!+\!\lambda_{k}\!-r})
     \big] \nn\\
  &&=\prod_{l=1}^{N}\prod_{m=1}^{l}\!\!
     \prod_{r=1}^{\lambda_{l}-\lambda_{l+1}} \!\!\!\!
     \big[ 
     \lambda_{m}\!+\!2(N\!-\!m)a\!
     +\!(1\!+\!(-1)^{\lambda_{m}\!-r})b\!-\!r\!+\!1 
     \big] \nn\\
  & &\quad\times\!\!\!\prod_{k=l+1}^{N}\!\!\!
     \big[ 
     (\lambda_{m}\!-\!\lambda_{k}\!+\!2(k-m)a\!-\!r\!+\!1)^{2}  \nn\\
  & &\hspace{20mm} 
     \!-\!2a^{2}(1\!-\!(-1)^{\lambda_{m}\!+\!\lambda_{k}\!-r})
     \big],
\end{eqnarray}
where the norm of the ground state is
\begin{eqnarray}
  \langle 1,1 \rangle
  &=&\frac{1}{\omega^{N(N-1)a+N(b+\frac{1}{2})}} \nn\\
  & &\times\prod_{j=1}^{N}
     \frac{\Gamma((j-1)a+b+\frac{1}{2})\Gamma(ja+1)}{\Gamma(1+a)},\nn
\end{eqnarray}
with $\Gamma(z)$ being the gamma function.
To get the norm of the general eigenstate 
with the composition $\lambda_{\sigma}$,
we recursively use the first relation of eqs.~\myref{eq:property-X-op_2}.
For example, in the case of $\sigma=(j,j+1)$, 
we perform the calculation,
\begin{eqnarray}
  \label{eq:norm-Laguerre2}
  &&\langle X_{j,j+1}\tilde{l}_{\lambda},
            X_{j,j+1}\tilde{l}_{\lambda}\rangle \nn\\
  &&=\!\big[(\bar{\lambda}_{j}\!-\!\bar{\lambda}_{j+1})^{2}
          \!-\! 2a^{2}(1\!+\!(-1)^{\lambda_{j}+\lambda_{j+1}})\big]
     \langle \tilde{l}_{\lambda},
             \tilde{l}_{\lambda} \rangle. \nn
\end{eqnarray}
Thus we can recursively calculate the norms for all the eigenstates.

%%%%%%%%%%%%%%%%%%%%%%%%%%%%%%%%%%%%%%%%%%%%%%%%%%%%%%%%%%%%%%%%%%%%%%%%%%%%
\section{Concluding Remarks}
\label{sec:concluding-remarks}
Generalizing the Takamura-Takano method,
we have investigated the $B_{N}$-Calogero model
with distinguishable particles and algebraically constructed
the nonsymmetric eigenfunctions for the model.
In other words, 
we have presented the Rodrigues formula for the nonsymmetric 
multivariable Laguerre polynomial.
The formula provides us with not only
known eigenfunctions which are even in all variables, 
but also those which have different parities for each variable.

So far, only nonsymmetric multivariable Laguerre polynomials
with even parities for all variables have been considered.
Under the restriction on parities,
the Cherednik operators for the $B_{N}$-Calogero model can be 
mapped to those for the $A_{N-1}$-Sutherland 
model~\cite{Baker_1,Baker_2,Baker_3,Kakei}.
That is, only the nonsymmetric multivariable Laguerre polynomials
which can be mapped to the well-investigated 
($A_{N-1}$-)Jack polynomials have been studied.
In this paper, we apply an algebraic method to the whole
nonsymmetric multivariable Laguerre polynomials with no restriction
on parities, which cannot generally be mapped to the nonsymmetric 
($A_{N-1}$-)Jack polynomials, for the first time.

The Calogero model with distinguishable particles 
we have considered in this paper
is mapped to the spin Calogero model 
by replacing the coordinate exchange operator $K_{jk}$
with the spin exchange operator $P_{jk}$
and identifying the reflection operators as unity.
This mapping is possible when the whole eigenfunction
is symmetric (or anti-symmetric) under the exchange of particles,
i.e.\ $K_{jk}P_{jk}=1\, (\mbox{or } -1)$,
and when the parities for each coordinate variable are even.
The eigenfunction we have constructed corresponds to 
the orbital part of the whole one.
To obtain the whole eigenfunction of the spin Calogero model,
we have to take an appropriate linear combination of products
of the orbital and spin parts.
We note that the particle exchange operators $K_{jk}P_{jk}$
and the reflection operators $t_{j}$
do not generally commute, $t_{j}K_{jk}P_{jk}=K_{jk}P_{jk}t_{k}$,
which is the reason why the nonsymmetric multivariable Laguerre
polynomials with different parities for each coordinate variable
have not been studied.
However the abandoned eigenstate is necessary
when we solve the diagonalization problem
for a more general $B_{N}$-Calogero model
explicitly including the reflection operators in the Hamiltonian.

In the limit $\omega\rightarrow\infty$,
the nonsymmetric multivariable Laguerre polynomial reduces to
the nonsymmetric $B_{N}$-Jack polynomial discussed 
in previous papers~\cite{Ujino_4,Ujino_5}.
We should note that the Rodrigues formula and an algebraic calculation
of the norm of the nonsymmetric $B_{N}$-Jack polynomials are
given in a parallel way to those of the nonsymmetric multivariable 
Laguerre polynomial.
The nonsymmetric $B_{N}$-Jack polynomials 
diagonalize the Cherednik operators 
\[
 D_{j}=x_{j}\nabla_{j}
 +a\!\!\!\sum_{k=j+1}^{N}\!\!\!(1+t_{j}t_{k})K_{jk}\!+\!b\,t_{j},
\]
whose physical significance is not clear at the moment.

An algebraic approach is very useful for construction of eigenstates
and calculation of physical quantities,
because of the connection with the concept of field theory.
It is interesting to verify whether
the symmetrized Takamura-Takano raising operators 
agree with the ones constructed 
by the Lapointe-Vinet method~\cite{Lapointe_1,Lapointe_2},
which is left for future study.

%%%%%%%%%%%%%%%%%%%%%%%%%%%%%%%%%%%%%%%%%%%%%%%%%%%%%%%%%%%%%%%%%%%%%%%%%%%%
\acknowledgement
The authors are grateful to Dr.~A.~Takamura for sending us the 
paper~\cite{Takamura} prior to publication.

One of the authors (HU) appreciates the Research Fellowships of the
Japan Society for the Promotion of Science for Young Scientists.

%%%%%%%%%%%%%%%%%%%%%%%%%%%%%%%%%%%%%%%%%%%%%%%%%%%%%%%%%%%%%%%%%%%%%%%%%%%%

\end{document}